\documentclass[aps,pra,twocolumn,superscriptaddress,showpacs]{revtex4}

\usepackage{graphicx}
\usepackage{natbib}
\usepackage[breaklinks]{hyperref}
\usepackage{amsmath}
\usepackage{mathtools}

\newcommand{\hflev}[4]{\textit{#1}$_{#2/#3}$, \textit{F}=#4}
\begin{document}


\title{Quantifying the role of thermal motion in free-space light-atom interaction}
\author{Yue-Sum Chin}
\affiliation{Center for Quantum Technologies, 3 Science Drive 2, Singapore 117543}
\author{Matthias Steiner}
\affiliation{Center for Quantum Technologies, 3 Science Drive 2, Singapore 117543}
\affiliation{Department of Physics, National University of Singapore, 2 Science Drive 3, Singapore 117542}
\author{Christian Kurtsiefer}
\affiliation{Center for Quantum Technologies, 3 Science Drive 2, Singapore 117543}
\affiliation{Department of Physics, National University of Singapore, 2 Science Drive 3, Singapore 117542}
\email[]{christian.kurtsiefer@gmail.com}
\date{\today}

\begin{abstract}
We demonstrate 17.7(1)\% extinction of a weak coherent field by a single atom. 
We observe a shift of the resonance frequency and a decrease in interaction strength with the external field when the atom, initially at 21(1)\,$\mu$K, is heated by the recoil of the scattered photons. 
Comparing to a simple model, we conclude that the initial temperature reduces the interaction strength by less than 10\%.
\end{abstract}

\pacs{
 32.90.+a,        
37.10.Gh, 
 37.10.Vz,	
 42.50.Ct      
 }

\maketitle
\section{Introduction}
The prospects of distributed quantum networks have triggered much effort in developing interfaces between single photons and single atoms (or other quantum emitters)~\cite{Kimble2008}. 
A major challenge lies in increasing the interaction strength of the atom
with incoming photons, which is a key ingredient for efficient transfer of quantum information from photons to atoms. 
While cavity-QED experiments have
 made  tremendous progress in this direction~\cite{Volz2011,Reiserer1349}, it remains an open question whether \mbox{(near-)deterministic} absorption of single photons is also possible without a cavity~\cite{Piro2011,Sandoghdar:2012,Victor2016,Brito:2016}. 

Single trapped atoms are a particularly good experimental platform for
quantitative comparisons of light-matter experiments with quantum optics
theory. 
The clean energy level structure and the trapping in ultra-high vacuum permits
deriving the interaction strength with a minimum of assumptions.
In a free space light-atom interface (as opposed to a situation with light
fields in cavities with a discrete mode spectrum), the interaction strength is characterized by a single parameter, the spatial mode overlap~$\Lambda\in [0,1]$, which quantifies the similarity of the incident light field to the atomic dipole mode~\cite{Golla2012,Sondermann2013}. 
The development of focusing schemes with large spatial mode overlap is a long-standing theoretical~\cite{Enk2000,Enk2004,Sondermann2007,Tey2009,Hetet2010} and experimental challenge~\cite{Wineland:87,Vamivakas2007,Gerhardt2007,Tey:2008,Wrigge2008,Aljunid2009,Piro2011,Pototschnig2011,Fischer:2014,Tran2016}. 
Approaches with multi-element
objectives~\cite{Vamivakas2007,Gerhardt2007,Piro2011,Tran2016},
singlet~\cite{Sortais2007,Tey:2008} and Fresnel lenses~\cite{Streed2011}, and
parabolic mirrors~\cite{Maiwald:2012,Alber2016} have been used with various single
emitter systems. 
However, the interaction strengths observed with these configurations~\cite{Fischer:2014,Tey2009} have fallen short of their theoretically expected capabilities.
Consequently, a better understanding of the underlying reasons  is necessary to further improve the interaction strength. 
Aside from imperfections of the focusing devices, the finite  positional
spread of the single atomic emitter is commonly suspected to reduce the interaction~\cite{Guthohrlein2001}.

In this paper, we present a light-atom interface based on a high numerical aperture lens and quantify the effect of insufficient localization of the atom on the light-atom interaction.
Initially at sub-Doppler temperatures, we heat the atom in a well-controlled
manner by scattering near-resonant photons and obtain a temperature
dependency of the interaction strength and resonance frequency.

This paper is organized as follows. 
In Sec.~\ref{sec:setup}, we describe the optical setup and the measurement sequence.  
We then characterize the light-atom interaction strength by a  transmission~(Sec.~\ref{sec:tx}) and a reflection~(Sec.~\ref{sec:sat}) measurement and
present the dependence of the light-atom interaction on
the positional spread of the atom in Sec.~\ref{sec:temp}.
\begin{figure} 
\centering
  \includegraphics[width=\columnwidth]{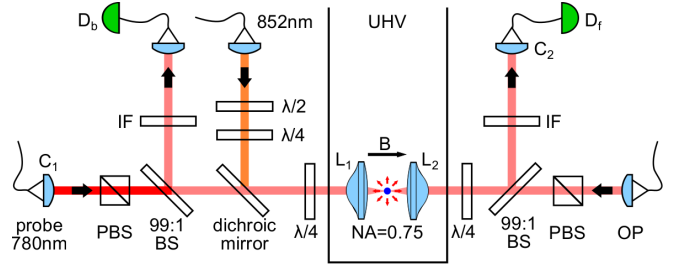}
  \caption{\label{fig:setup} Setup for probing light-atom interaction in free space. 
  D:~detector, UHV:~ultra-high vacuum chamber, 
  IF:~interference filter centered at 780\,nm, $\lambda$/2:~half-wave plate, $\lambda$/4:~quarter-wave plate, 
 C:~fiber coupling lens, PBS:~polarizing beam splitter, BS:~beam splitter, L:~high numerical aperture lens, B:~magnetic field, OP: optical pumping.
}
\end{figure}
\section{Experimental setup and measurement sequence}\label{sec:setup}
The core of the optical setup is a pair of high numerical
aperture lenses $L_1$ and~$L_2$ (NA=0.75, focal length~$f$=5.95\,mm, see Fig.~\ref{fig:setup}). 
A single $^{87}$Rb atom is trapped at the joint focus of these lenses with a
far-off-resonant, red detuned optical dipole trap~(852\,nm)~\cite{Schlosser2001,Schlosser2002}.
The circularly polarized ($\sigma^+$) trap has a depth of
$U_0=k_\textrm{B}\times2.22(1)$\,mK, with measured radial frequencies
$\omega_x/2\pi=107(1)$\,kHz and $\omega_y/2\pi=124(1)$\,kHz, and an axial
frequency $\omega_z/2\pi=13.8(1)$\,kHz. 

We probe the light-atom interaction by driving the closed transition \mbox{5\hflev{S}{1}{2}{2}, $m_F$=-2} to  \mbox{5\hflev{P}{3}{2}{3}, $m_F$=-3} near 780\,nm. 
The spatial mode of the incident probe field is defined by the aperture of the
single mode fiber, the collimation lens~$C_1$, and the focusing lens~$L_1$.  
The beam profile before $L_1$ is approximately Gaussian, with a
waist~$w_L=2.7$\,mm.
Following~\cite{Tey2009,Aljunid2013}, the spatial mode overlap~$\Lambda$ of
the Gaussian mode focused by an ideal lens with the dipole mode of a stationary atom depends on the focusing strength~$u \vcentcolon = w_L/f$,
\begin{equation}\label{eq:lambda}
\Lambda= \frac{3}{16 u^3} e^{2/u^2} \left[ \Gamma\left(-\frac{1}{4},\frac{1}{u^2} \right)  + u \Gamma\left(\frac{1}{4},\frac{1}{u^2} \right)\right]^2,
\end{equation}
where $\Gamma(a,b)$ is the incomplete gamma function. For our experimental parameters, we expect~$\Lambda=11.2\%$.

\begin{figure}
\centering
  \includegraphics[width=\columnwidth]{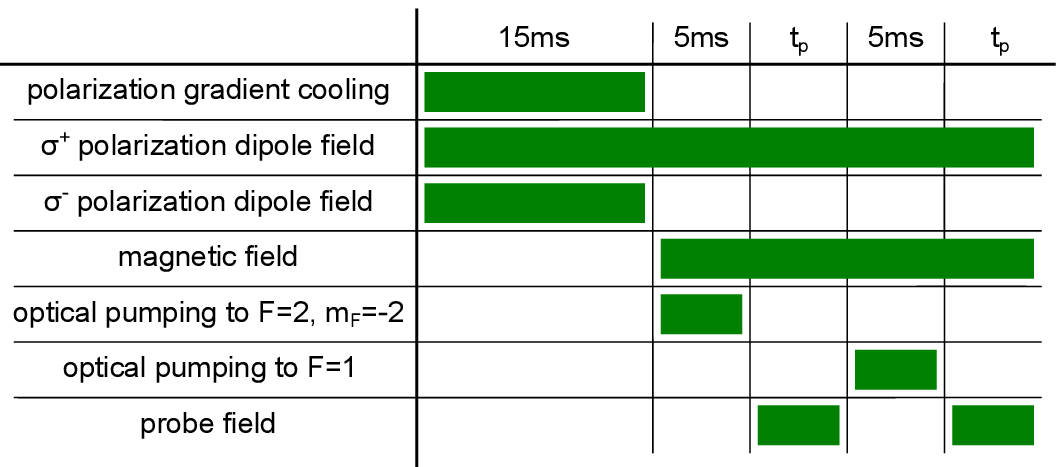}
  \caption{\label{fig:sequence} Experimental sequence to probe the light-atom
    interaction. 
}
\end{figure}
The experimental sequence used in Sec.~\ref{sec:tx}, \ref{sec:sat}, and~\ref{sec:temp} is depicted in Fig.~\ref{fig:sequence}.    
After loading a single atom into the dipole trap, the atom is cooled by
polarization gradient cooling (PGC)~\cite{Lett1988}.  
For efficient cooling, we apply an additional $\sigma^-$-polarized dipole field~(852\,nm) injected through the same optical fiber as the $\sigma^+$-polarized dipole field. 
The  $\sigma^-$-polarized dipole field, which  is switched off after the PGC phase, originates from an independent laser running several hundreds of GHz detuned from the $\sigma^+$-polarized dipole field. 
Subsequently, a bias magnetic field of~$0.74$\,mT is applied along the optical axis, and the atom is prepared in the 5\hflev{S}{1}{2}{2}, $m_F$=-2 state by optical pumping. 
Next, the probe field is switched on for a duration~$t_\textrm{p}$ during which the detection events at avalanche photodetectors (APD) $D_\textrm{b}$ and $D_\textrm{f}$ are recorded. 
Finally, we perform a reference measurement to determine the power of the probe pulse. 
Optically pumping to the \mbox{5\hflev{S}{1}{2}{1}} hyperfine state shifts the atom out of resonance with the probe field by 6.8\,GHz. 
The probe pulse is reapplied for a time~$t_\textrm{p}$, and we infer the average number of incident probe photons at the position of the atom from counts at detector~$D_\textrm{f}$ during the reference pulse, taking into account the optical losses from the position of the atom to detector~$D_\textrm{f}$. 

We determine the detection efficiencies of 
$D_\textrm{b}$ and~$D_\textrm{f}$ by comparing  against a
calibrated pin photodiode and a calibrated APD to $\eta_\textrm{b}=59(3)\,\%$ and $\eta_\textrm{f}=56(4)\,\%$, respectively.
The experimental detection rates presented in the following 
are background-corrected for 300~cps at detector~$D_\textrm{b}$ and 155~cps at detector~$D_\textrm{f}$.  

\section{Extinction measurement}\label{sec:tx}
In this section, we describe an extinction measurement to determine the spatial mode overlap~$\Lambda$ between probe and atomic dipole mode. 
For this, we compare the transmitted power through the system during the probe and the reference phase.
To detect the transmitted power,  the probe mode is re-collimated by the
second aspheric lens~$L_2$ and then coupled into a single mode fiber
directing the light to the forward detector~$D_\textrm{f}$. 
The total electric field $\vec{E'}(\vec{r})$ of the light moving away from the
atom is a superposition of the probe field~$\vec{E}_\textrm{p}(\vec{r})$ and
the field scattered by the atom~$\vec{E}_\textrm{sc}(\vec{r})$:
\begin{equation}
\vec{E'}(\vec{r})= \vec{E}_\textrm{p}(\vec{r})+ \vec{E}_\textrm{sc}(\vec{r})\,.
\end{equation}
The electric field amplitude~$E_\textrm{f}=\int \vec{E'}(\vec{r})G^*(\vec{r})
dS$ at the detector~$D_\textrm{f}$ is given by the spatial mode overlap  of the total electric field with the collection mode~$G(\vec{r})$ ($dS$ is a differential area element perpendicular
to the optical axis)~\cite{Aljunid2009}. 
In this configuration,
$\Lambda$ cannot be deduced from the transmitted power without knowledge or assumptions about this mode overlap~\cite{Wineland:87,Gerhardt2007,Vamivakas2007,Hwang2007,Wrigge2008,Tey:2008}. 
The relative transmission~$\tau\left(\omega_\textrm{p}\right)$, which is the optical power at detector~$D_\textrm{f}$ normalized to the reference power, contains Lorentzian and dispersion-like terms~\cite{Gerhardt2007},
\begin{align}\label{eq:tx}
\tau\left(\omega_\textrm{p}\right) =& 1  + A^2 \mathcal{L}\left(\omega_\textrm{p}\right) \nonumber\\
&+2A  \mathcal{L}\left(\omega_\textrm{p}\right) \left[ \left(\omega_\textrm{p}-\omega_0-\delta\omega\right) \sin{\phi} - \frac{\Gamma}{2}\cos{\phi} \right]\,,
\end{align}
where
$\mathcal{L}\left(\omega_\textrm{p}\right)=1/\left[\left(\omega_\textrm{p}-\omega_0-\delta\omega\right)^2+\Gamma^2/4\right]$
is a Lorentzian profile with linewidth ~$\Gamma$, $\omega_\textrm{p}$ is the frequency of the probe field, and coefficient~$A$ and the phase $\phi$  depend on the mode matching of the probe and the collection mode.
The resonance frequency shift $\delta\omega= \omega_{z} + \omega_\textrm{ac}$
from the natural transition frequency~$\omega_0$  is due to a Zeeman shift
$\omega_\textrm{z}$ and an AC Stark shift $\omega_\textrm{ac}$. 
For perfect mode matching (e.g. when the collimation lens is identical to the focusing lens), the coefficients in Eq.~(\ref{eq:tx}) simplify to $A=\Gamma \Lambda$ and $\phi=0$. 
The transmission spectrum takes a purely Lorentzian form with a resonant extinction~$\epsilon = 4\Lambda \left( 1-\Lambda\right)$~\cite{Aljunid2009}. 

\begin{figure}
\centering
  \includegraphics[width=\columnwidth]{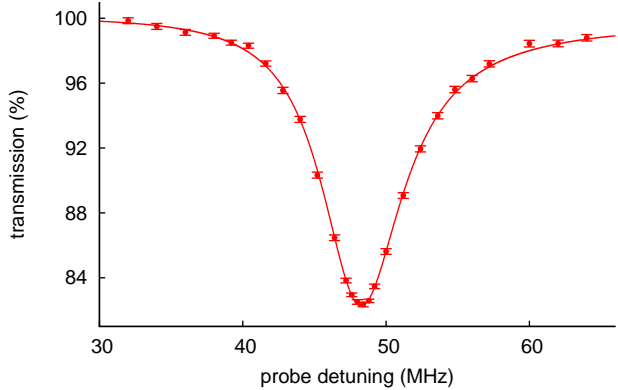}
  \caption{\label{fig:tx}
    Transmission measurement of a weak coherent probe beam.
    The solid line is a fit of Eq.~(\ref{eq:tx}) with free parameters:
    linewidth~$\Gamma/2\pi=6.9(1)$\,MHz, frequency shift~~$\delta\omega=48.03(3)\,$MHz, spatial
    overlap~$\Lambda=4.67(2)\,\%$, and phase~$\phi_0=0.13(1)\,$rad ($\chi^2_{\text{red}}=1.01$),
    resulting in a resonant extinction of $\epsilon=17.7(1)\%$.
   Error bars represent one standard deviation due to propagated Poissonian
   counting uncertainties.
}
\end{figure}

We measure the transmission of a weak probe field for $t_\textrm{p}=20$\,ms containing on average 550\,photons per pulse.
Tuning the frequency of the probe field, we find a maximum extinction~$\epsilon$~= 17.7(1)\%~(Fig.~\ref{fig:tx}).
The observed transmission spectrum shows a small deviation from a Lorentzian profile. 
This deviation is caused by the imperfect mode overlap between probe and collection mode. 
We infer a mode overlap of approximately $70\%$ from the probe power measured at detector~$D_\textrm{f}$, corrected for losses of the optical elements. 
To account for the small deviation from the ideal case, we include the phase~$\phi$ as a free fit parameter. 
The model in Eq.~(\ref{eq:tx}) fits the observed values with four free parameters
($\chi^2_{\text{red}}=1.01$): frequency shift~$\delta\omega=48.03(3)\,$MHz, spatial overlap~$\Lambda=4.67(2)\,\%$, phase~$\phi_0=0.13(1)\,$rad, and linewidth~$\Gamma/2\pi=6.9(1)$\,MHz (slightly broader
than the natural linewidth~$\Gamma_0/2\pi=6.07$\,MHz~\cite{Volz1996}). 
This interaction strength is 50\% larger compared to our previous experiments
with lenses of smaller numerical aperture (NA=$0.55$, \cite{Tey:2008}).

\section{Saturation measurement}\label{sec:sat} 
\begin{figure}
\centering
  \includegraphics[width=\columnwidth]{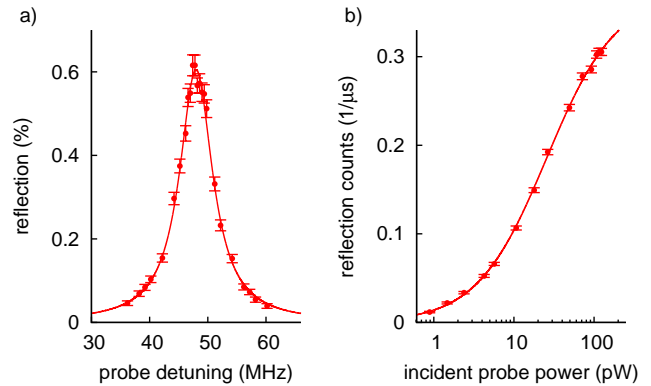}
  \caption{\label{fig:refl_sat}
  (a) Light scattered into the backward detector~$D_\textrm{b}$ for different
  probe detunings. The solid line is a Lorentzian fit of 
  Eq.~(\ref{eq:refl}) with free parameters linewidth~$\Gamma/2\pi=6.9(1)$\,MHz,
  frequency shift~$\delta\omega/2\pi= 48.0(1)$\,MHz, and resonant backscattering
  probability~$P_\textrm{b,0}=0.61(1)\%$, with $\chi^2_{\text{red}}=1.03$.
  (b)~Resonant saturation measurement, with the solid line representing the fit
  of Eq.~(\ref{eq:sat}) with saturation power~$P_{\textrm{sat}}=26(2)$\,pW and
  total detection efficiency~~$\eta=1.95(2)$\% as free parameters ($\chi^2_{\text{red}}=1.3$). Error bars represent one standard deviation due to
  propagated Poissonian counting uncertainties.  
}
\end{figure}

We also determine~$\Lambda$ from the intensity of the atomic fluorescence at backward detector~$D_\textrm{b}$.
Figure~\ref{fig:refl_sat}(a) shows the probability~$P_\textrm{b}$ for an
incident photon to be backscattered by the atom when tuning the frequency~$\omega_\textrm{p}$ of the probe field. 
This value is obtained by normalizing the number of detected photons at detector~$D_\textrm{b}$ to the average number of incident photons  during the  probe phase~$t_\textrm{p}=20$\,$\mu \text{s}$~\cite{Syed2011,Agio:2008}.  
The backscattering probability is proportional to the atomic excited state population and therefore follows a Lorentzian profile
\begin{equation}\label{eq:refl}
 P_\textrm{b} =  \frac{P_\textrm{b,0}}{4\left(\omega_\textrm{p}-\omega_0-\delta\omega\right)^2/\Gamma^2+1} \,,
\end{equation}
where $P_\textrm{b,0}$ is the {\em resonant} backscattering probability. 
The experimental values of $P_{\text{b}}$ in Fig.~\ref{fig:refl_sat} can be well
described by this model, with a frequency
shift~$\delta\omega/2\pi= 48.0(1)$\,MHz from the natural transition frequency,
$P_\textrm{b,0}=0.61(1)\%$, and $\Gamma/2\pi=6.9(1)$\,MHz. 

The incident power needed to saturate the target transition is a direct measurement of~$\Lambda$.
For a resonantly driven two-level atom, the saturation power~$P_{\textrm{sat}}$ is  given by
\begin{equation}\label{eq:sat_power}
P_{\textrm{sat}} = \frac{\hbar \omega_0 \Gamma_0}{8} \frac{1}{\Lambda}\,,
\end{equation}
where $\omega_0 $ is the transition frequency~\cite{Fischer:2014}. 
For complete mode matching ($\Lambda=1$), Eq.~(\ref{eq:sat_power}) gives a saturation power~$P_{\textrm{sat},\Lambda=1}=1.21$\,pW for the considered transition. 
The spatial overlap~$\Lambda = P_{\textrm{sat}}/P_{\textrm{sat},\Lambda=1}$ is obtained from the experimentally determined saturation power~$P_{\textrm{sat}}$. 

The saturation  power~$P_{\textrm{sat}}$ is determined by varying the
excitation power on resonance [see Fig.~\ref{fig:refl_sat}(b)]. 
We use a short probe interval ($t_\textrm{p}=4$\,$\mu \text{s}$) to minimize heating of the
atom. 
A saturation power of $P_{\textrm{sat}}=26(2)$\,pW and a total detection
efficiency~$\eta=1.95(2)$\% are obtained from fitting the resultant atomic
fluorescence rate $R_\textrm{b}$ to the expected saturation function
\begin{equation}\label{eq:sat}
 R_\textrm{b} = \frac{\eta \Gamma_0}{2} \frac{P_{\textrm{inc}}}{P_{\textrm{inc}} + P_\textrm{sat}}\,,
\end{equation}
where~$ P_\textrm{inc}$ is the power of the incident beam at the position of the atom.
We infer a total collection
efficiency~$\eta_\textrm{sm}=\eta/\eta_\textrm{b}=3.3(3)\%$ into a single mode
fiber, which is compatible with the highest
efficiencies reported for free space optic~\cite{Hucul2015,Ghadimi2016}. 
Comparing $P_{\textrm{sat}}$ to $P_{\textrm{sat},\Lambda=1}$ indicates a spatial overlap~$\Lambda = 4.7(4)\%$, in agreement with the extinction measurement~$\Lambda=4.67(2)\,\%$.  
The uncertainty of the spatial overlap is dominated by the uncertainty of the efficiency~$\eta_\textrm{f}$ of detector~$D_\textrm{f}$, which we use in conjunction with a set of calibrated neutral density filters  to determine the incident power~$ P_\textrm{inc}$. 

\begin{figure}
\centering
  \includegraphics[width=\columnwidth]{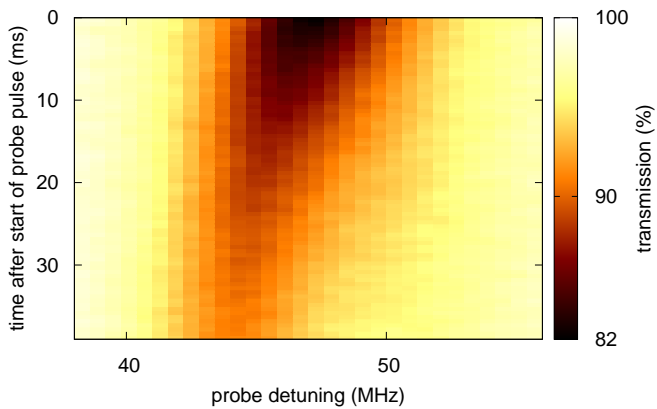}
  \caption{\label{fig:tx_matrix}
Time-resolved extinction measurement. Each row presents a transmission
spectrum similar to Fig.~\ref{fig:tx} and is obtained by collecting
photodetection events in 0.5\,ms wide time bins.
As the atom is heated by scattering probe photons, the transmission increases, and also the
frequency of the minimal transmission shifts to a lower detuning
from the unperturbed resonance.
}
\end{figure}

\section{Temperature dependence of light-atom interaction}\label{sec:temp}
We investigate whether the residual temperature of the atom limits the coupling to the probe field.
As the recoil associated with the scattering of the probe field increases the kinetic energy of the atom, different atom temperatures can be accessed by following the temporal evolution of the probe transmission. 
The photodetection events during the probe phase are time-tagged and sorted
into 0.5\,ms wide time bins, resulting in the time-resolved transmission
spectrum shown in Fig.~\ref{fig:tx_matrix}.
The probe pulse has a length of $t_\textrm{p}=40$\,ms and contains on average about 9000 photons.
As the probe pulse progresses, the resonance frequency shifts towards lower frequencies, and the extinction reduces.  

\begin{figure}
\centering
  \includegraphics[width=\columnwidth]{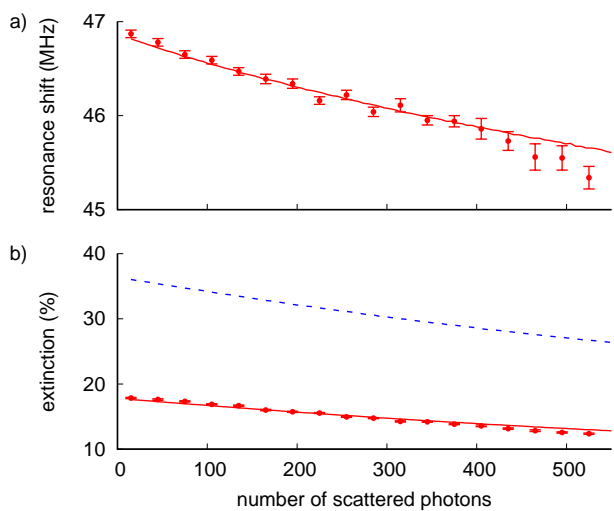}
  \caption{\label{fig:scat_tx_res}
  The effect of recoil heating on the resonance frequency~(a) and extinction~(b)  obtained by rearranging the histogram in Fig.~\ref{fig:tx_matrix} with a bin width of 30 scattered photons. 
  Resonance frequency  and extinction decreases fairly linearly as the atom heats up. 
  (a)~Solid red line is the numerical result of Eq.~(\ref{eq:tx_T}) with the
  frequency shift at the center of trap~$\delta\omega(0)$ as a free fit
  parameter ($\chi^2_{\text{red}}=1.4$). 
  (b)~Dashed blue line is the numerical result of Eq.~(\ref{eq:tx_T}) without free fit parameter. 
  The relative temperature dependence is well reproduced by assuming an
  effective interaction strength~$\Lambda_\textrm{eff}(\vec{r})=
  \left(1-\alpha\right)\Lambda(\vec{r})$, fit parameter
  $\alpha=0.54(1)$~(solid red line, $\chi^2_{\text{red}}=11.6)$. 
  Error bars represent one standard deviation obtained from least-squares fit of the individual spectra.
  }
\end{figure}

For a quantitative analysis, we sort the recorded relative transmission according to the number of scattered photons for each probe frequency. 
The time-integrated number of scattered photons~$n_\textrm{s}(t) $ from the
beginning of the probe phase until the time $t \in [0,t_p]$ is deduced from
the number of detected photons $n_{p}(t)$ and $n_\textrm{ref}(t)$ at detector~$D_\textrm{f}$ during the  probe
 and the reference phase, taking into account the optical losses from the atom to the detector~$\eta_\textrm{op}=59(5)\%$ and the detection efficiency~$\eta_\textrm{f}$,  
\begin{equation}
n_\textrm{s}(t) = \left[n_\textrm{ref}(t)-n_\textrm{p}(t) \right]/\eta_\textrm{f} \eta_\textrm{op}\,. 
\end{equation}
We choose a relative bin width of 30 scattered photons and obtain the resonance frequency and the extinction by fitting to Eq.~(\ref{eq:tx}). 
The resonance frequency and the extinction decrease fairly linearly with the number of scattered photons~(Fig.~\ref{fig:scat_tx_res}). 
After scattering approximately 500 photons, the resonance frequency is lowered by~$1.5(1)$\,MHz, and the extinction is reduced by approximately~$30\,\%$ to~$\epsilon=12.4(1)$\%. 

We derive the temperature dependent transmission spectrum by including the
spatial dependence of the frequency shift~$\delta\omega(\vec{r}) = \omega_{z}
+ \omega_\textrm{ac}(\vec{r})$ and the mode
overlap~$\Lambda(\vec{r})$~\cite{Teo2011} in Eq.~(\ref{eq:tx}), where $\vec{r}$ is the position of the atom relative to the centre of the trap. 
The AC Stark shift~$\omega_\textrm{ac}(\vec{r})$ is treated in the paraxial approximation, given the large beam waist of~$1.4\,\mu$m of the dipole trap. 
For the tightly focused probe field, we evaluate the spatial dependence of
the mode overlap~$\Lambda(\vec{r})$ according to \cite{Tey2009}. 
The transmission spectrum, averaged over many different spatial configurations,
is then given by
\begin{equation}\label{eq:tx_T}
 \langle\tau\rangle = \int  p(T,\vec{r})\,\tau(\vec{r}) d^3r\,,
\end{equation}
where~$p(T,\vec{r})$ is the probability distribution of the atom position. 
We treat the motion of the atom classically and assume that the probability distribution~$p(T,\vec{r})$ is
governed by a Maxwell-Boltzmann distribution with standard deviations of the positional spread
of the atom~$\sigma_i=\sqrt{k_\textrm{B}
  T/mw_\textrm{i}^2}$, with $i=x,y,z$ and mass $m$ of $^{87}$Rb. 
Equation~(\ref{eq:tx_T}) can then be evaluated by a Monte-Carlo method.  
Each scattered photon increases the total energy of the atom  by $2E_\textrm{r}$, where $E_\textrm{r}= \hbar^2 k^2/2m$ is the photon recoil energy. 
The gained energy is anisotropically distributed because of the uni-directional excitation by the probe beam.  
Each photon leads therefore, on average, to an energy increase of~$\frac{2}{3}E_\textrm{r}$ in the radial directions, and $\frac{4}{3}E_\textrm{r}$ in the axial direction. 
From a release-recapture technique~\cite{PhysRevA.78.033425},  we infer an initial atom temperature of~$21(1)\,\mu$K.
Thus, after 500 scattering events the axial temperature is increased by approximately 120\,$\mu$K to just below  Doppler temperature $T_\textrm{D}=146\,\mu$K. 

The frequency shift expected from Eq.~(\ref{eq:tx_T}) matches well with the
experimental results~[(Fig.~\ref{fig:scat_tx_res}(a)], where we use only the
frequency shift at the center of the trap~$\delta\omega(0)=47.32(5)\,$MHz as a free fit parameter.
This good agreement indicates that the model captures the effect of the dipole trap well. 
The initial resonance frequency  is slightly lower compared to the results in Sec.~\ref{sec:sat} and~\ref{sec:tx} because of a slightly lower dipole trap power. 
Figure~\ref{fig:scat_tx_res}(b) (dashed blue line) shows the theoretical extinction expected from Eq.~(\ref{eq:tx_T}) with our focusing parameters. 
A large discrepancy between experiment and theory in the absolute magnitude of the extinction is evident. 
However, the relative magnitude of the reduction of the extinction as a function of scattered photons is well reproduced by Eq.~(\ref{eq:tx_T}).  
To illustrate this agreement, we heuristically introduce an effective interaction strength~$\Lambda_\textrm{eff}(\vec{r})= \left(1-\alpha\right)\Lambda(\vec{r})$ where $\alpha$ describes the  reduction of the interaction strength. 
Including the reduction~$\alpha=0.54(1)$ as a free parameter and fit to
Eq.~(\ref{eq:tx_T}), the temperature dependence of the extinction is closely
matched (see Fig.~\ref{fig:scat_tx_res}(b), solid red line).
Using the effective interaction we extrapolate a spatial overlap~$\Lambda=5.1\%$ for a stationary atom which is approximately 10\% larger than the interaction observed for our lowest temperatures. 
This estimation provides an upper bound for the temperature effect because our model treats the atomic motion classically and therefore does not include the finite spread of the motional ground state.  
Thus, we can exclude the initial atom temperature as the main reason for the large discrepancy to the ideal theoretical expectation. 
Imperfections of the focusing lens and deviations of the incident field from a Gaussian beam are likely to cause the lower interaction strength. 

Finally, we discuss possible origins of the observed linewidth broadening~(Fig.~\ref{fig:tx} and~\ref{fig:refl_sat}). 
Doppler and power broadening are negligible because of the low atomic temperature of 21(1)\,$\mu$K and the weak excitation field in both measurements~$P_\textrm{probe}< 0.02 P_\textrm{sat}$. 
We use Eq.~(\ref{eq:tx_T}) to estimate whether the broadening is caused by the thermal motion in the spatially varying trap potential. 
We find an expected linewidth of~$6.3$\,MHz for~$T=21$\,$\mu$K.  
Therefore, we attribute the residual linewidth broadening to other noise sources, such as the linewidth of the probe laser. 

\section{Conclusion}\label{sec:con}
We demonstrated an effective spatial mode overlap~$\Lambda=4.7(4)\%$ between an external probe mode and the atomic dipole mode, and showed that the light-atom interaction can be limited by the residual motion of the atom even at sub-Doppler temperatures. 
The spatially varying AC Stark shift and the tight confinement of the probe field cause a reduction of approximately 10\% in interaction strength for our lowest atom temperatures. 
Thus, further cooling  to the motional ground state promises only a moderate improvement~\cite{Kaufman2012,Thompson2013}. 
The relatively small effect of the thermal motion of the atom hints to imperfections of the focusing lens as main cause for the low interaction compared to the tight focusing theory outlined in~\cite{Tey2009}. 
Significant improvement of the interaction strength requires therefore a
careful analysis of the focusing lens and the  application of aberration
corrections to the incident probe field.
Coherent control of the atomic motion and temporal shaping of the incoming photon can further optimize the absorption efficiency~\cite{Trautmann2016,Victor2016}.

\begin{acknowledgments}
We thank V.\,Leong and N.\,Lewty for contributions in an early stage of the experiment. 
We acknowledge the support of this work by the Ministry of Education in
Singapore (AcRF Tier 1) and the National Research Foundation, Prime Minister's
office.
M.\,Steiner acknowledges support by the Lee Kuan Yew Postdoctoral Fellowship.
\end{acknowledgments}
\bibliographystyle{apsrev4-1}

\begin{thebibliography}{43}%
\makeatletter
\providecommand \@ifxundefined [1]{%
 \@ifx{#1\undefined}
}%
\providecommand \@ifnum [1]{%
 \ifnum #1\expandafter \@firstoftwo
 \else \expandafter \@secondoftwo
 \fi
}%
\providecommand \@ifx [1]{%
 \ifx #1\expandafter \@firstoftwo
 \else \expandafter \@secondoftwo
 \fi
}%
\providecommand \natexlab [1]{#1}%
\providecommand \enquote  [1]{``#1''}%
\providecommand \bibnamefont  [1]{#1}%
\providecommand \bibfnamefont [1]{#1}%
\providecommand \citenamefont [1]{#1}%
\providecommand \href@noop [0]{\@secondoftwo}%
\providecommand \href [0]{\begingroup \@sanitize@url \@href}%
\providecommand \@href[1]{\@@startlink{#1}\@@href}%
\providecommand \@@href[1]{\endgroup#1\@@endlink}%
\providecommand \@sanitize@url [0]{\catcode `\\12\catcode `\$12\catcode
  `\&12\catcode `\#12\catcode `\^12\catcode `\_12\catcode `\%12\relax}%
\providecommand \@@startlink[1]{}%
\providecommand \@@endlink[0]{}%
\providecommand \url  [0]{\begingroup\@sanitize@url \@url }%
\providecommand \@url [1]{\endgroup\@href {#1}{\urlprefix }}%
\providecommand \urlprefix  [0]{URL }%
\providecommand \Eprint [0]{\href }%
\providecommand \doibase [0]{http://dx.doi.org/}%
\providecommand \selectlanguage [0]{\@gobble}%
\providecommand \bibinfo  [0]{\@secondoftwo}%
\providecommand \bibfield  [0]{\@secondoftwo}%
\providecommand \translation [1]{[#1]}%
\providecommand \BibitemOpen [0]{}%
\providecommand \bibitemStop [0]{}%
\providecommand \bibitemNoStop [0]{.\EOS\space}%
\providecommand \EOS [0]{\spacefactor3000\relax}%
\providecommand \BibitemShut  [1]{\csname bibitem#1\endcsname}%
\let\auto@bib@innerbib\@empty
\bibitem [{\citenamefont {Kimble}(2008)}]{Kimble2008}%
  \BibitemOpen
  \bibfield  {author} {\bibinfo {author} {\bibfnamefont {H.~J.}\ \bibnamefont
  {Kimble}},\ }\href {http://dx.doi.org/10.1038/nature07127} {\bibfield
  {journal} {\bibinfo  {journal} {Nature}\ }\textbf {\bibinfo {volume} {453}},\
  \bibinfo {pages} {1023} (\bibinfo {year} {2008})}\BibitemShut {NoStop}%
\bibitem [{\citenamefont {Volz}\ \emph {et~al.}(2011)\citenamefont {Volz},
  \citenamefont {Gehr}, \citenamefont {Dubois}, \citenamefont {Esteve},\ and\
  \citenamefont {Reichel}}]{Volz2011}%
  \BibitemOpen
  \bibfield  {author} {\bibinfo {author} {\bibfnamefont {J.}~\bibnamefont
  {Volz}}, \bibinfo {author} {\bibfnamefont {R.}~\bibnamefont {Gehr}}, \bibinfo
  {author} {\bibfnamefont {G.}~\bibnamefont {Dubois}}, \bibinfo {author}
  {\bibfnamefont {J.}~\bibnamefont {Esteve}}, \ and\ \bibinfo {author}
  {\bibfnamefont {J.}~\bibnamefont {Reichel}},\ }\href
  {http://dx.doi.org/10.1038/nature10225} {\bibfield  {journal} {\bibinfo
  {journal} {Nature}\ }\textbf {\bibinfo {volume} {475}},\ \bibinfo {pages}
  {210} (\bibinfo {year} {2011})}\BibitemShut {NoStop}%
\bibitem [{\citenamefont {Reiserer}\ \emph {et~al.}(2013)\citenamefont
  {Reiserer}, \citenamefont {Ritter},\ and\ \citenamefont
  {Rempe}}]{Reiserer1349}%
  \BibitemOpen
  \bibfield  {author} {\bibinfo {author} {\bibfnamefont {A.}~\bibnamefont
  {Reiserer}}, \bibinfo {author} {\bibfnamefont {S.}~\bibnamefont {Ritter}}, \
  and\ \bibinfo {author} {\bibfnamefont {G.}~\bibnamefont {Rempe}},\ }\href
  {\doibase 10.1126/science.1246164} {\bibfield  {journal} {\bibinfo  {journal}
  {Science}\ }\textbf {\bibinfo {volume} {342}},\ \bibinfo {pages} {1349}
  (\bibinfo {year} {2013})}\BibitemShut {NoStop}%
\bibitem [{\citenamefont {Piro}\ \emph {et~al.}(2011)\citenamefont {Piro},
  \citenamefont {Rohde}, \citenamefont {Schuck}, \citenamefont {Almendros},
  \citenamefont {Huwer}, \citenamefont {Ghosh}, \citenamefont {Haase},
  \citenamefont {Hennrich}, \citenamefont {Dubin},\ and\ \citenamefont
  {Eschner}}]{Piro2011}%
  \BibitemOpen
  \bibfield  {author} {\bibinfo {author} {\bibfnamefont {N.}~\bibnamefont
  {Piro}}, \bibinfo {author} {\bibfnamefont {F.}~\bibnamefont {Rohde}},
  \bibinfo {author} {\bibfnamefont {C.}~\bibnamefont {Schuck}}, \bibinfo
  {author} {\bibfnamefont {M.}~\bibnamefont {Almendros}}, \bibinfo {author}
  {\bibfnamefont {J.}~\bibnamefont {Huwer}}, \bibinfo {author} {\bibfnamefont
  {J.}~\bibnamefont {Ghosh}}, \bibinfo {author} {\bibfnamefont
  {A.}~\bibnamefont {Haase}}, \bibinfo {author} {\bibfnamefont
  {M.}~\bibnamefont {Hennrich}}, \bibinfo {author} {\bibfnamefont
  {F.}~\bibnamefont {Dubin}}, \ and\ \bibinfo {author} {\bibfnamefont
  {J.}~\bibnamefont {Eschner}},\ }\href {http://dx.doi.org/10.1038/nphys1805}
  {\bibfield  {journal} {\bibinfo  {journal} {Nat Phys}\ }\textbf {\bibinfo
  {volume} {7}},\ \bibinfo {pages} {17} (\bibinfo {year} {2011})}\BibitemShut
  {NoStop}%
\bibitem [{\citenamefont {Rezus}\ \emph {et~al.}(2012)\citenamefont {Rezus},
  \citenamefont {Walt}, \citenamefont {Lettow}, \citenamefont {Renn},
  \citenamefont {Zumofen}, \citenamefont {G\"otzinger},\ and\ \citenamefont
  {Sandoghdar}}]{Sandoghdar:2012}%
  \BibitemOpen
  \bibfield  {author} {\bibinfo {author} {\bibfnamefont {Y.~L.~A.}\
  \bibnamefont {Rezus}}, \bibinfo {author} {\bibfnamefont {S.~G.}\ \bibnamefont
  {Walt}}, \bibinfo {author} {\bibfnamefont {R.}~\bibnamefont {Lettow}},
  \bibinfo {author} {\bibfnamefont {A.}~\bibnamefont {Renn}}, \bibinfo {author}
  {\bibfnamefont {G.}~\bibnamefont {Zumofen}}, \bibinfo {author} {\bibfnamefont
  {S.}~\bibnamefont {G\"otzinger}}, \ and\ \bibinfo {author} {\bibfnamefont
  {V.}~\bibnamefont {Sandoghdar}},\ }\href {\doibase
  10.1103/PhysRevLett.108.093601} {\bibfield  {journal} {\bibinfo  {journal}
  {Phys. Rev. Lett.}\ }\textbf {\bibinfo {volume} {108}},\ \bibinfo {pages}
  {093601} (\bibinfo {year} {2012})}\BibitemShut {NoStop}%
\bibitem [{\citenamefont {Leong}\ \emph {et~al.}(2016)\citenamefont {Leong},
  \citenamefont {Seidler}, \citenamefont {Steiner}, \citenamefont {Cer\`e},\
  and\ \citenamefont {Kurtsiefer}}]{Victor2016}%
  \BibitemOpen
  \bibfield  {author} {\bibinfo {author} {\bibfnamefont {V.}~\bibnamefont
  {Leong}}, \bibinfo {author} {\bibfnamefont {M.~A.}\ \bibnamefont {Seidler}},
  \bibinfo {author} {\bibfnamefont {M.}~\bibnamefont {Steiner}}, \bibinfo
  {author} {\bibfnamefont {A.}~\bibnamefont {Cer\`e}}, \ and\ \bibinfo {author}
  {\bibfnamefont {C.}~\bibnamefont {Kurtsiefer}},\ }\href@noop {} {\bibfield
  {journal} {\bibinfo  {journal} {arxiv.org}\ ,\ \bibinfo {pages}
  {arXiv:1604.08020}} (\bibinfo {year} {2016})}\BibitemShut {NoStop}%
\bibitem [{\citenamefont {Brito}\ \emph {et~al.}(2016)\citenamefont {Brito},
  \citenamefont {Kucera}, \citenamefont {Eich}, \citenamefont {M{\"u}ller},\
  and\ \citenamefont {Eschner}}]{Brito:2016}%
  \BibitemOpen
  \bibfield  {author} {\bibinfo {author} {\bibfnamefont {J.}~\bibnamefont
  {Brito}}, \bibinfo {author} {\bibfnamefont {S.}~\bibnamefont {Kucera}},
  \bibinfo {author} {\bibfnamefont {P.}~\bibnamefont {Eich}}, \bibinfo {author}
  {\bibfnamefont {P.}~\bibnamefont {M{\"u}ller}}, \ and\ \bibinfo {author}
  {\bibfnamefont {J.}~\bibnamefont {Eschner}},\ }\href {\doibase
  10.1007/s00340-015-6276-9} {\bibfield  {journal} {\bibinfo  {journal}
  {Applied Physics B}\ }\textbf {\bibinfo {volume} {122}},\ \bibinfo {pages}
  {1} (\bibinfo {year} {2016})}\BibitemShut {NoStop}%
\bibitem [{\citenamefont {Golla}\ \emph {et~al.}(2012)\citenamefont {Golla},
  \citenamefont {Chalopin}, \citenamefont {Bader}, \citenamefont {Harder},
  \citenamefont {Mantel}, \citenamefont {Maiwald}, \citenamefont {Lindlein},
  \citenamefont {Sondermann},\ and\ \citenamefont {Leuchs}}]{Golla2012}%
  \BibitemOpen
  \bibfield  {author} {\bibinfo {author} {\bibfnamefont {A.}~\bibnamefont
  {Golla}}, \bibinfo {author} {\bibfnamefont {B.}~\bibnamefont {Chalopin}},
  \bibinfo {author} {\bibfnamefont {M.}~\bibnamefont {Bader}}, \bibinfo
  {author} {\bibfnamefont {I.}~\bibnamefont {Harder}}, \bibinfo {author}
  {\bibfnamefont {K.}~\bibnamefont {Mantel}}, \bibinfo {author} {\bibfnamefont
  {R.}~\bibnamefont {Maiwald}}, \bibinfo {author} {\bibfnamefont
  {N.}~\bibnamefont {Lindlein}}, \bibinfo {author} {\bibfnamefont
  {M.}~\bibnamefont {Sondermann}}, \ and\ \bibinfo {author} {\bibfnamefont
  {G.}~\bibnamefont {Leuchs}},\ }\href {\doibase 10.1140/epjd/e2012-30293-y}
  {\bibfield  {journal} {\bibinfo  {journal} {The European Physical Journal D}\
  }\textbf {\bibinfo {volume} {66}},\ \bibinfo {pages} {1} (\bibinfo {year}
  {2012})}\BibitemShut {NoStop}%
\bibitem [{\citenamefont {Leuchs}\ and\ \citenamefont
  {Sondermann}(2013)}]{Sondermann2013}%
  \BibitemOpen
  \bibfield  {author} {\bibinfo {author} {\bibfnamefont {G.}~\bibnamefont
  {Leuchs}}\ and\ \bibinfo {author} {\bibfnamefont {M.}~\bibnamefont
  {Sondermann}},\ }\href {\doibase 10.1080/09500340.2012.716461} {\bibfield
  {journal} {\bibinfo  {journal} {Journal of Modern Optics}\ }\textbf {\bibinfo
  {volume} {60}},\ \bibinfo {pages} {36} (\bibinfo {year} {2013})}\BibitemShut
  {NoStop}%
\bibitem [{\citenamefont {van Enk}\ and\ \citenamefont
  {Kimble}(2000)}]{Enk2000}%
  \BibitemOpen
  \bibfield  {author} {\bibinfo {author} {\bibfnamefont {S.~J.}\ \bibnamefont
  {van Enk}}\ and\ \bibinfo {author} {\bibfnamefont {H.~J.}\ \bibnamefont
  {Kimble}},\ }\href {\doibase 10.1103/PhysRevA.61.051802} {\bibfield
  {journal} {\bibinfo  {journal} {Phys. Rev. A}\ }\textbf {\bibinfo {volume}
  {61}},\ \bibinfo {pages} {051802} (\bibinfo {year} {2000})}\BibitemShut
  {NoStop}%
\bibitem [{\citenamefont {van Enk}(2004)}]{Enk2004}%
  \BibitemOpen
  \bibfield  {author} {\bibinfo {author} {\bibfnamefont {S.~J.}\ \bibnamefont
  {van Enk}},\ }\href {\doibase 10.1103/PhysRevA.69.043813} {\bibfield
  {journal} {\bibinfo  {journal} {Phys. Rev. A}\ }\textbf {\bibinfo {volume}
  {69}},\ \bibinfo {pages} {043813} (\bibinfo {year} {2004})}\BibitemShut
  {NoStop}%
\bibitem [{\citenamefont {Sondermann}\ \emph {et~al.}(2007)\citenamefont
  {Sondermann}, \citenamefont {Maiwald}, \citenamefont {Konermann},
  \citenamefont {Lindlein}, \citenamefont {Peschel},\ and\ \citenamefont
  {Leuchs}}]{Sondermann2007}%
  \BibitemOpen
  \bibfield  {author} {\bibinfo {author} {\bibfnamefont {M.}~\bibnamefont
  {Sondermann}}, \bibinfo {author} {\bibfnamefont {R.}~\bibnamefont {Maiwald}},
  \bibinfo {author} {\bibfnamefont {H.}~\bibnamefont {Konermann}}, \bibinfo
  {author} {\bibfnamefont {N.}~\bibnamefont {Lindlein}}, \bibinfo {author}
  {\bibfnamefont {U.}~\bibnamefont {Peschel}}, \ and\ \bibinfo {author}
  {\bibfnamefont {G.}~\bibnamefont {Leuchs}},\ }\href {\doibase
  10.1007/s00340-007-2859-4} {\bibfield  {journal} {\bibinfo  {journal}
  {Applied Physics B}\ }\textbf {\bibinfo {volume} {89}},\ \bibinfo {pages}
  {489} (\bibinfo {year} {2007})}\BibitemShut {NoStop}%
\bibitem [{\citenamefont {Tey}\ \emph {et~al.}(2009)\citenamefont {Tey},
  \citenamefont {Maslennikov}, \citenamefont {Liew}, \citenamefont {Aljunid},
  \citenamefont {Huber}, \citenamefont {Chng}, \citenamefont {Chen},
  \citenamefont {Scarani},\ and\ \citenamefont {Kurtsiefer}}]{Tey2009}%
  \BibitemOpen
  \bibfield  {author} {\bibinfo {author} {\bibfnamefont {M.~K.}\ \bibnamefont
  {Tey}}, \bibinfo {author} {\bibfnamefont {G.}~\bibnamefont {Maslennikov}},
  \bibinfo {author} {\bibfnamefont {T.~C.~H.}\ \bibnamefont {Liew}}, \bibinfo
  {author} {\bibfnamefont {S.~A.}\ \bibnamefont {Aljunid}}, \bibinfo {author}
  {\bibfnamefont {F.}~\bibnamefont {Huber}}, \bibinfo {author} {\bibfnamefont
  {B.}~\bibnamefont {Chng}}, \bibinfo {author} {\bibfnamefont {Z.}~\bibnamefont
  {Chen}}, \bibinfo {author} {\bibfnamefont {V.}~\bibnamefont {Scarani}}, \
  and\ \bibinfo {author} {\bibfnamefont {C.}~\bibnamefont {Kurtsiefer}},\
  }\href {http://stacks.iop.org/1367-2630/11/i=4/a=043011} {\bibfield
  {journal} {\bibinfo  {journal} {New Journal of Physics}\ }\textbf {\bibinfo
  {volume} {11}},\ \bibinfo {pages} {043011} (\bibinfo {year}
  {2009})}\BibitemShut {NoStop}%
\bibitem [{\citenamefont {H\'etet}\ \emph {et~al.}(2010)\citenamefont
  {H\'etet}, \citenamefont {Slodi\ifmmode~\check{c}\else \v{c}\fi{}ka},
  \citenamefont {Gl\"atzle}, \citenamefont {Hennrich},\ and\ \citenamefont
  {Blatt}}]{Hetet2010}%
  \BibitemOpen
  \bibfield  {author} {\bibinfo {author} {\bibfnamefont {G.}~\bibnamefont
  {H\'etet}}, \bibinfo {author} {\bibfnamefont {L.}~\bibnamefont
  {Slodi\ifmmode~\check{c}\else \v{c}\fi{}ka}}, \bibinfo {author}
  {\bibfnamefont {A.}~\bibnamefont {Gl\"atzle}}, \bibinfo {author}
  {\bibfnamefont {M.}~\bibnamefont {Hennrich}}, \ and\ \bibinfo {author}
  {\bibfnamefont {R.}~\bibnamefont {Blatt}},\ }\href {\doibase
  10.1103/PhysRevA.82.063812} {\bibfield  {journal} {\bibinfo  {journal} {Phys.
  Rev. A}\ }\textbf {\bibinfo {volume} {82}},\ \bibinfo {pages} {063812}
  (\bibinfo {year} {2010})}\BibitemShut {NoStop}%
\bibitem [{\citenamefont {Wineland}\ \emph {et~al.}(1987)\citenamefont
  {Wineland}, \citenamefont {Itano},\ and\ \citenamefont
  {Bergquist}}]{Wineland:87}%
  \BibitemOpen
  \bibfield  {author} {\bibinfo {author} {\bibfnamefont {D.~J.}\ \bibnamefont
  {Wineland}}, \bibinfo {author} {\bibfnamefont {W.~M.}\ \bibnamefont {Itano}},
  \ and\ \bibinfo {author} {\bibfnamefont {J.~C.}\ \bibnamefont {Bergquist}},\
  }\href {\doibase 10.1364/OL.12.000389} {\bibfield  {journal} {\bibinfo
  {journal} {Opt. Lett.}\ }\textbf {\bibinfo {volume} {12}},\ \bibinfo {pages}
  {389} (\bibinfo {year} {1987})}\BibitemShut {NoStop}%
\bibitem [{\citenamefont {Vamivakas}\ \emph {et~al.}(2007)\citenamefont
  {Vamivakas}, \citenamefont {Atat\"ure}, \citenamefont {Dreiser},
  \citenamefont {Yilmaz}, \citenamefont {Badolato}, \citenamefont {Swan},
  \citenamefont {Goldberg}, \citenamefont {Imamo{\u{g}}lu},\ and\ \citenamefont
  {\"Unl\"u}}]{Vamivakas2007}%
  \BibitemOpen
  \bibfield  {author} {\bibinfo {author} {\bibfnamefont {A.~N.}\ \bibnamefont
  {Vamivakas}}, \bibinfo {author} {\bibfnamefont {M.}~\bibnamefont
  {Atat\"ure}}, \bibinfo {author} {\bibfnamefont {J.}~\bibnamefont {Dreiser}},
  \bibinfo {author} {\bibfnamefont {S.~T.}\ \bibnamefont {Yilmaz}}, \bibinfo
  {author} {\bibfnamefont {A.}~\bibnamefont {Badolato}}, \bibinfo {author}
  {\bibfnamefont {A.~K.}\ \bibnamefont {Swan}}, \bibinfo {author}
  {\bibfnamefont {B.~B.}\ \bibnamefont {Goldberg}}, \bibinfo {author}
  {\bibfnamefont {A.}~\bibnamefont {Imamo{\u{g}}lu}}, \ and\ \bibinfo {author}
  {\bibfnamefont {M.~S.}\ \bibnamefont {\"Unl\"u}},\ }\href {\doibase
  10.1021/nl0717255} {\bibfield  {journal} {\bibinfo  {journal} {Nano Letters}\
  }\textbf {\bibinfo {volume} {7}},\ \bibinfo {pages} {2892} (\bibinfo {year}
  {2007})}\BibitemShut {NoStop}%
\bibitem [{\citenamefont {Gerhardt}\ \emph {et~al.}(2007)\citenamefont
  {Gerhardt}, \citenamefont {Wrigge}, \citenamefont {Bushev}, \citenamefont
  {Zumofen}, \citenamefont {Agio}, \citenamefont {Pfab},\ and\ \citenamefont
  {Sandoghdar}}]{Gerhardt2007}%
  \BibitemOpen
  \bibfield  {author} {\bibinfo {author} {\bibfnamefont {I.}~\bibnamefont
  {Gerhardt}}, \bibinfo {author} {\bibfnamefont {G.}~\bibnamefont {Wrigge}},
  \bibinfo {author} {\bibfnamefont {P.}~\bibnamefont {Bushev}}, \bibinfo
  {author} {\bibfnamefont {G.}~\bibnamefont {Zumofen}}, \bibinfo {author}
  {\bibfnamefont {M.}~\bibnamefont {Agio}}, \bibinfo {author} {\bibfnamefont
  {R.}~\bibnamefont {Pfab}}, \ and\ \bibinfo {author} {\bibfnamefont
  {V.}~\bibnamefont {Sandoghdar}},\ }\href {\doibase
  10.1103/PhysRevLett.98.033601} {\bibfield  {journal} {\bibinfo  {journal}
  {Phys. Rev. Lett.}\ }\textbf {\bibinfo {volume} {98}},\ \bibinfo {pages}
  {033601} (\bibinfo {year} {2007})}\BibitemShut {NoStop}%
\bibitem [{\citenamefont {Tey}\ \emph {et~al.}(2008)\citenamefont {Tey},
  \citenamefont {Chen}, \citenamefont {Aljunid}, \citenamefont {Chng},
  \citenamefont {Huber}, \citenamefont {Maslennikov},\ and\ \citenamefont
  {Kurtsiefer}}]{Tey:2008}%
  \BibitemOpen
  \bibfield  {author} {\bibinfo {author} {\bibfnamefont {M.~K.}\ \bibnamefont
  {Tey}}, \bibinfo {author} {\bibfnamefont {Z.}~\bibnamefont {Chen}}, \bibinfo
  {author} {\bibfnamefont {S.~A.}\ \bibnamefont {Aljunid}}, \bibinfo {author}
  {\bibfnamefont {B.}~\bibnamefont {Chng}}, \bibinfo {author} {\bibfnamefont
  {F.}~\bibnamefont {Huber}}, \bibinfo {author} {\bibfnamefont
  {G.}~\bibnamefont {Maslennikov}}, \ and\ \bibinfo {author} {\bibfnamefont
  {C.}~\bibnamefont {Kurtsiefer}},\ }\href@noop {} {\bibfield  {journal}
  {\bibinfo  {journal} {Nature Physics}\ }\textbf {\bibinfo {volume} {4}},\
  \bibinfo {pages} {924} (\bibinfo {year} {2008})}\BibitemShut {NoStop}%
\bibitem [{\citenamefont {Wrigge}\ \emph {et~al.}(2008)\citenamefont {Wrigge},
  \citenamefont {Gerhardt}, \citenamefont {Hwang}, \citenamefont {Zumofen},\
  and\ \citenamefont {Sandoghdar}}]{Wrigge2008}%
  \BibitemOpen
  \bibfield  {author} {\bibinfo {author} {\bibfnamefont {G.}~\bibnamefont
  {Wrigge}}, \bibinfo {author} {\bibfnamefont {I.}~\bibnamefont {Gerhardt}},
  \bibinfo {author} {\bibfnamefont {J.}~\bibnamefont {Hwang}}, \bibinfo
  {author} {\bibfnamefont {G.}~\bibnamefont {Zumofen}}, \ and\ \bibinfo
  {author} {\bibfnamefont {V.}~\bibnamefont {Sandoghdar}},\ }\href
  {http://dx.doi.org/10.1038/nphys812} {\bibfield  {journal} {\bibinfo
  {journal} {Nat Phys}\ }\textbf {\bibinfo {volume} {4}},\ \bibinfo {pages}
  {60} (\bibinfo {year} {2008})}\BibitemShut {NoStop}%
\bibitem [{\citenamefont {Aljunid}\ \emph {et~al.}(2009)\citenamefont
  {Aljunid}, \citenamefont {Tey}, \citenamefont {Chng}, \citenamefont {Liew},
  \citenamefont {Maslennikov}, \citenamefont {Scarani},\ and\ \citenamefont
  {Kurtsiefer}}]{Aljunid2009}%
  \BibitemOpen
  \bibfield  {author} {\bibinfo {author} {\bibfnamefont {S.~A.}\ \bibnamefont
  {Aljunid}}, \bibinfo {author} {\bibfnamefont {M.~K.}\ \bibnamefont {Tey}},
  \bibinfo {author} {\bibfnamefont {B.}~\bibnamefont {Chng}}, \bibinfo {author}
  {\bibfnamefont {T.}~\bibnamefont {Liew}}, \bibinfo {author} {\bibfnamefont
  {G.}~\bibnamefont {Maslennikov}}, \bibinfo {author} {\bibfnamefont
  {V.}~\bibnamefont {Scarani}}, \ and\ \bibinfo {author} {\bibfnamefont
  {C.}~\bibnamefont {Kurtsiefer}},\ }\href {\doibase
  10.1103/PhysRevLett.103.153601} {\bibfield  {journal} {\bibinfo  {journal}
  {Phys. Rev. Lett.}\ }\textbf {\bibinfo {volume} {103}},\ \bibinfo {pages}
  {153601} (\bibinfo {year} {2009})}\BibitemShut {NoStop}%
\bibitem [{\citenamefont {Pototschnig}\ \emph {et~al.}(2011)\citenamefont
  {Pototschnig}, \citenamefont {Chassagneux}, \citenamefont {Hwang},
  \citenamefont {Zumofen}, \citenamefont {Renn},\ and\ \citenamefont
  {Sandoghdar}}]{Pototschnig2011}%
  \BibitemOpen
  \bibfield  {author} {\bibinfo {author} {\bibfnamefont {M.}~\bibnamefont
  {Pototschnig}}, \bibinfo {author} {\bibfnamefont {Y.}~\bibnamefont
  {Chassagneux}}, \bibinfo {author} {\bibfnamefont {J.}~\bibnamefont {Hwang}},
  \bibinfo {author} {\bibfnamefont {G.}~\bibnamefont {Zumofen}}, \bibinfo
  {author} {\bibfnamefont {A.}~\bibnamefont {Renn}}, \ and\ \bibinfo {author}
  {\bibfnamefont {V.}~\bibnamefont {Sandoghdar}},\ }\href {\doibase
  10.1103/PhysRevLett.107.063001} {\bibfield  {journal} {\bibinfo  {journal}
  {Phys. Rev. Lett.}\ }\textbf {\bibinfo {volume} {107}},\ \bibinfo {pages}
  {063001} (\bibinfo {year} {2011})}\BibitemShut {NoStop}%
\bibitem [{\citenamefont {Fischer}\ \emph {et~al.}(2014)\citenamefont
  {Fischer}, \citenamefont {Bader}, \citenamefont {Maiwald}, \citenamefont
  {Golla}, \citenamefont {Sondermann},\ and\ \citenamefont
  {Leuchs}}]{Fischer:2014}%
  \BibitemOpen
  \bibfield  {author} {\bibinfo {author} {\bibfnamefont {M.}~\bibnamefont
  {Fischer}}, \bibinfo {author} {\bibfnamefont {M.}~\bibnamefont {Bader}},
  \bibinfo {author} {\bibfnamefont {R.}~\bibnamefont {Maiwald}}, \bibinfo
  {author} {\bibfnamefont {A.}~\bibnamefont {Golla}}, \bibinfo {author}
  {\bibfnamefont {M.}~\bibnamefont {Sondermann}}, \ and\ \bibinfo {author}
  {\bibfnamefont {G.}~\bibnamefont {Leuchs}},\ }\href {\doibase
  10.1007/s00340-014-5817-y} {\bibfield  {journal} {\bibinfo  {journal}
  {Applied Physics B}\ }\textbf {\bibinfo {volume} {117}},\ \bibinfo {pages}
  {797} (\bibinfo {year} {2014})}\BibitemShut {NoStop}%
\bibitem [{\citenamefont {Tran}\ \emph {et~al.}(2016)\citenamefont {Tran},
  \citenamefont {Wrachtrup},\ and\ \citenamefont {Gerhardt}}]{Tran2016}%
  \BibitemOpen
  \bibfield  {author} {\bibinfo {author} {\bibfnamefont {T.~H.}\ \bibnamefont
  {Tran}}, \bibinfo {author} {\bibfnamefont {J.}~\bibnamefont {Wrachtrup}}, \
  and\ \bibinfo {author} {\bibfnamefont {I.}~\bibnamefont {Gerhardt}},\
  }\href@noop {} {\bibfield  {journal} {\bibinfo  {journal} {arxiv.org}\ ,\
  \bibinfo {pages} {arXiv:1608.05224}} (\bibinfo {year} {2016})}\BibitemShut
  {NoStop}%
\bibitem [{\citenamefont {Sortais}\ \emph {et~al.}(2007)\citenamefont
  {Sortais}, \citenamefont {Marion}, \citenamefont {Tuchendler}, \citenamefont
  {Lance}, \citenamefont {Lamare}, \citenamefont {Fournet}, \citenamefont
  {Armellin}, \citenamefont {Mercier}, \citenamefont {Messin}, \citenamefont
  {Browaeys},\ and\ \citenamefont {Grangier}}]{Sortais2007}%
  \BibitemOpen
  \bibfield  {author} {\bibinfo {author} {\bibfnamefont {Y.~R.~P.}\
  \bibnamefont {Sortais}}, \bibinfo {author} {\bibfnamefont {H.}~\bibnamefont
  {Marion}}, \bibinfo {author} {\bibfnamefont {C.}~\bibnamefont {Tuchendler}},
  \bibinfo {author} {\bibfnamefont {A.~M.}\ \bibnamefont {Lance}}, \bibinfo
  {author} {\bibfnamefont {M.}~\bibnamefont {Lamare}}, \bibinfo {author}
  {\bibfnamefont {P.}~\bibnamefont {Fournet}}, \bibinfo {author} {\bibfnamefont
  {C.}~\bibnamefont {Armellin}}, \bibinfo {author} {\bibfnamefont
  {R.}~\bibnamefont {Mercier}}, \bibinfo {author} {\bibfnamefont
  {G.}~\bibnamefont {Messin}}, \bibinfo {author} {\bibfnamefont
  {A.}~\bibnamefont {Browaeys}}, \ and\ \bibinfo {author} {\bibfnamefont
  {P.}~\bibnamefont {Grangier}},\ }\href {\doibase 10.1103/PhysRevA.75.013406}
  {\bibfield  {journal} {\bibinfo  {journal} {Phys. Rev. A}\ }\textbf {\bibinfo
  {volume} {75}},\ \bibinfo {pages} {013406} (\bibinfo {year}
  {2007})}\BibitemShut {NoStop}%
\bibitem [{\citenamefont {Streed}\ \emph {et~al.}(2011)\citenamefont {Streed},
  \citenamefont {Norton}, \citenamefont {Jechow}, \citenamefont {Weinhold},\
  and\ \citenamefont {Kielpinski}}]{Streed2011}%
  \BibitemOpen
  \bibfield  {author} {\bibinfo {author} {\bibfnamefont {E.~W.}\ \bibnamefont
  {Streed}}, \bibinfo {author} {\bibfnamefont {B.~G.}\ \bibnamefont {Norton}},
  \bibinfo {author} {\bibfnamefont {A.}~\bibnamefont {Jechow}}, \bibinfo
  {author} {\bibfnamefont {T.~J.}\ \bibnamefont {Weinhold}}, \ and\ \bibinfo
  {author} {\bibfnamefont {D.}~\bibnamefont {Kielpinski}},\ }\href {\doibase
  10.1103/PhysRevLett.106.010502} {\bibfield  {journal} {\bibinfo  {journal}
  {Phys. Rev. Lett.}\ }\textbf {\bibinfo {volume} {106}},\ \bibinfo {pages}
  {010502} (\bibinfo {year} {2011})}\BibitemShut {NoStop}%
\bibitem [{\citenamefont {Maiwald}\ \emph {et~al.}(2012)\citenamefont
  {Maiwald}, \citenamefont {Golla}, \citenamefont {Fischer}, \citenamefont
  {Bader}, \citenamefont {Heugel}, \citenamefont {Chalopin}, \citenamefont
  {Sondermann},\ and\ \citenamefont {Leuchs}}]{Maiwald:2012}%
  \BibitemOpen
  \bibfield  {author} {\bibinfo {author} {\bibfnamefont {R.}~\bibnamefont
  {Maiwald}}, \bibinfo {author} {\bibfnamefont {A.}~\bibnamefont {Golla}},
  \bibinfo {author} {\bibfnamefont {M.}~\bibnamefont {Fischer}}, \bibinfo
  {author} {\bibfnamefont {M.}~\bibnamefont {Bader}}, \bibinfo {author}
  {\bibfnamefont {S.}~\bibnamefont {Heugel}}, \bibinfo {author} {\bibfnamefont
  {B.}~\bibnamefont {Chalopin}}, \bibinfo {author} {\bibfnamefont
  {M.}~\bibnamefont {Sondermann}}, \ and\ \bibinfo {author} {\bibfnamefont
  {G.}~\bibnamefont {Leuchs}},\ }\href {\doibase 10.1103/PhysRevA.86.043431}
  {\bibfield  {journal} {\bibinfo  {journal} {Phys. Rev. A}\ }\textbf {\bibinfo
  {volume} {86}},\ \bibinfo {pages} {043431} (\bibinfo {year}
  {2012})}\BibitemShut {NoStop}%
\bibitem [{\citenamefont {Alber}\ \emph {et~al.}(2016)\citenamefont {Alber},
  \citenamefont {Fischer}, \citenamefont {Bader}, \citenamefont {Mantel},
  \citenamefont {Sondermann},\ and\ \citenamefont {Leuchs}}]{Alber2016}%
  \BibitemOpen
  \bibfield  {author} {\bibinfo {author} {\bibfnamefont {L.}~\bibnamefont
  {Alber}}, \bibinfo {author} {\bibfnamefont {M.}~\bibnamefont {Fischer}},
  \bibinfo {author} {\bibfnamefont {M.}~\bibnamefont {Bader}}, \bibinfo
  {author} {\bibfnamefont {K.}~\bibnamefont {Mantel}}, \bibinfo {author}
  {\bibfnamefont {M.}~\bibnamefont {Sondermann}}, \ and\ \bibinfo {author}
  {\bibfnamefont {G.}~\bibnamefont {Leuchs}},\ }\href@noop {} {\bibfield
  {journal} {\bibinfo  {journal} {arxiv.org}\ ,\ \bibinfo {pages}
  {arXiv:1609.06884}} (\bibinfo {year} {2016})}\BibitemShut {NoStop}%
\bibitem [{\citenamefont {Guthohrlein}\ \emph {et~al.}(2001)\citenamefont
  {Guthohrlein}, \citenamefont {Keller}, \citenamefont {Hayasaka},
  \citenamefont {Lange},\ and\ \citenamefont {Walther}}]{Guthohrlein2001}%
  \BibitemOpen
  \bibfield  {author} {\bibinfo {author} {\bibfnamefont {G.~R.}\ \bibnamefont
  {Guthohrlein}}, \bibinfo {author} {\bibfnamefont {M.}~\bibnamefont {Keller}},
  \bibinfo {author} {\bibfnamefont {K.}~\bibnamefont {Hayasaka}}, \bibinfo
  {author} {\bibfnamefont {W.}~\bibnamefont {Lange}}, \ and\ \bibinfo {author}
  {\bibfnamefont {H.}~\bibnamefont {Walther}},\ }\href
  {http://dx.doi.org/10.1038/35102129} {\bibfield  {journal} {\bibinfo
  {journal} {Nature}\ }\textbf {\bibinfo {volume} {414}},\ \bibinfo {pages}
  {49} (\bibinfo {year} {2001})}\BibitemShut {NoStop}%
\bibitem [{\citenamefont {Schlosser}\ \emph {et~al.}(2001)\citenamefont
  {Schlosser}, \citenamefont {Reymond}, \citenamefont {Protsenko},\ and\
  \citenamefont {Grangier}}]{Schlosser2001}%
  \BibitemOpen
  \bibfield  {author} {\bibinfo {author} {\bibfnamefont {N.}~\bibnamefont
  {Schlosser}}, \bibinfo {author} {\bibfnamefont {G.}~\bibnamefont {Reymond}},
  \bibinfo {author} {\bibfnamefont {I.}~\bibnamefont {Protsenko}}, \ and\
  \bibinfo {author} {\bibfnamefont {P.}~\bibnamefont {Grangier}},\ }\href
  {http://dx.doi.org/10.1038/35082512} {\bibfield  {journal} {\bibinfo
  {journal} {Nature}\ }\textbf {\bibinfo {volume} {411}},\ \bibinfo {pages}
  {1024} (\bibinfo {year} {2001})}\BibitemShut {NoStop}%
\bibitem [{\citenamefont {Schlosser}\ \emph {et~al.}(2002)\citenamefont
  {Schlosser}, \citenamefont {Reymond},\ and\ \citenamefont
  {Grangier}}]{Schlosser2002}%
  \BibitemOpen
  \bibfield  {author} {\bibinfo {author} {\bibfnamefont {N.}~\bibnamefont
  {Schlosser}}, \bibinfo {author} {\bibfnamefont {G.}~\bibnamefont {Reymond}},
  \ and\ \bibinfo {author} {\bibfnamefont {P.}~\bibnamefont {Grangier}},\
  }\href {\doibase 10.1103/PhysRevLett.89.023005} {\bibfield  {journal}
  {\bibinfo  {journal} {Phys. Rev. Lett.}\ }\textbf {\bibinfo {volume} {89}},\
  \bibinfo {pages} {023005} (\bibinfo {year} {2002})}\BibitemShut {NoStop}%
\bibitem [{\citenamefont {Aljunid}\ \emph {et~al.}(2013)\citenamefont
  {Aljunid}, \citenamefont {Maslennikov}, \citenamefont {Wang}, \citenamefont
  {Dao}, \citenamefont {Scarani},\ and\ \citenamefont
  {Kurtsiefer}}]{Aljunid2013}%
  \BibitemOpen
  \bibfield  {author} {\bibinfo {author} {\bibfnamefont {S.~A.}\ \bibnamefont
  {Aljunid}}, \bibinfo {author} {\bibfnamefont {G.}~\bibnamefont
  {Maslennikov}}, \bibinfo {author} {\bibfnamefont {Y.}~\bibnamefont {Wang}},
  \bibinfo {author} {\bibfnamefont {H.~L.}\ \bibnamefont {Dao}}, \bibinfo
  {author} {\bibfnamefont {V.}~\bibnamefont {Scarani}}, \ and\ \bibinfo
  {author} {\bibfnamefont {C.}~\bibnamefont {Kurtsiefer}},\ }\href {\doibase
  10.1103/PhysRevLett.111.103001} {\bibfield  {journal} {\bibinfo  {journal}
  {Phys. Rev. Lett.}\ }\textbf {\bibinfo {volume} {111}},\ \bibinfo {pages}
  {103001} (\bibinfo {year} {2013})}\BibitemShut {NoStop}%
\bibitem [{\citenamefont {Lett}\ \emph {et~al.}(1988)\citenamefont {Lett},
  \citenamefont {Watts}, \citenamefont {Westbrook}, \citenamefont {Phillips},
  \citenamefont {Gould},\ and\ \citenamefont {Metcalf}}]{Lett1988}%
  \BibitemOpen
  \bibfield  {author} {\bibinfo {author} {\bibfnamefont {P.~D.}\ \bibnamefont
  {Lett}}, \bibinfo {author} {\bibfnamefont {R.~N.}\ \bibnamefont {Watts}},
  \bibinfo {author} {\bibfnamefont {C.~I.}\ \bibnamefont {Westbrook}}, \bibinfo
  {author} {\bibfnamefont {W.~D.}\ \bibnamefont {Phillips}}, \bibinfo {author}
  {\bibfnamefont {P.~L.}\ \bibnamefont {Gould}}, \ and\ \bibinfo {author}
  {\bibfnamefont {H.~J.}\ \bibnamefont {Metcalf}},\ }\href {\doibase
  10.1103/PhysRevLett.61.169} {\bibfield  {journal} {\bibinfo  {journal} {Phys.
  Rev. Lett.}\ }\textbf {\bibinfo {volume} {61}},\ \bibinfo {pages} {169}
  (\bibinfo {year} {1988})}\BibitemShut {NoStop}%
\bibitem [{\citenamefont {Hwang}\ and\ \citenamefont
  {Moerner}(2007)}]{Hwang2007}%
  \BibitemOpen
  \bibfield  {author} {\bibinfo {author} {\bibfnamefont {J.}~\bibnamefont
  {Hwang}}\ and\ \bibinfo {author} {\bibfnamefont {W.}~\bibnamefont
  {Moerner}},\ }\href {\doibase http://dx.doi.org/10.1016/j.optcom.2007.08.032}
  {\bibfield  {journal} {\bibinfo  {journal} {Optics Communications}\ }\textbf
  {\bibinfo {volume} {280}},\ \bibinfo {pages} {487 } (\bibinfo {year}
  {2007})}\BibitemShut {NoStop}%
\bibitem [{\citenamefont {Volz}\ and\ \citenamefont
  {Schmoranzer}(1996)}]{Volz1996}%
  \BibitemOpen
  \bibfield  {author} {\bibinfo {author} {\bibfnamefont {U.}~\bibnamefont
  {Volz}}\ and\ \bibinfo {author} {\bibfnamefont {H.}~\bibnamefont
  {Schmoranzer}},\ }\href {http://stacks.iop.org/1402-4896/1996/i=T65/a=007}
  {\bibfield  {journal} {\bibinfo  {journal} {Physica Scripta}\ }\textbf
  {\bibinfo {volume} {1996}},\ \bibinfo {pages} {48} (\bibinfo {year}
  {1996})}\BibitemShut {NoStop}%
\bibitem [{\citenamefont {Aljunid}\ \emph {et~al.}(2011)\citenamefont
  {Aljunid}, \citenamefont {Chng}, \citenamefont {Lee}, \citenamefont
  {Paesold}, \citenamefont {Maslennikov},\ and\ \citenamefont
  {Kurtsiefer}}]{Syed2011}%
  \BibitemOpen
  \bibfield  {author} {\bibinfo {author} {\bibfnamefont {S.~A.}\ \bibnamefont
  {Aljunid}}, \bibinfo {author} {\bibfnamefont {B.}~\bibnamefont {Chng}},
  \bibinfo {author} {\bibfnamefont {J.}~\bibnamefont {Lee}}, \bibinfo {author}
  {\bibfnamefont {M.}~\bibnamefont {Paesold}}, \bibinfo {author} {\bibfnamefont
  {G.}~\bibnamefont {Maslennikov}}, \ and\ \bibinfo {author} {\bibfnamefont
  {C.}~\bibnamefont {Kurtsiefer}},\ }\href {\doibase
  10.1080/09500340.2010.522780} {\bibfield  {journal} {\bibinfo  {journal}
  {Journal of Modern Optics}\ }\textbf {\bibinfo {volume} {58}},\ \bibinfo
  {pages} {299} (\bibinfo {year} {2011})}\BibitemShut {NoStop}%
\bibitem [{\citenamefont {Zumofen}\ \emph {et~al.}(2008)\citenamefont
  {Zumofen}, \citenamefont {Mojarad}, \citenamefont {Sandoghdar},\ and\
  \citenamefont {Agio}}]{Agio:2008}%
  \BibitemOpen
  \bibfield  {author} {\bibinfo {author} {\bibfnamefont {G.}~\bibnamefont
  {Zumofen}}, \bibinfo {author} {\bibfnamefont {N.~M.}\ \bibnamefont
  {Mojarad}}, \bibinfo {author} {\bibfnamefont {V.}~\bibnamefont {Sandoghdar}},
  \ and\ \bibinfo {author} {\bibfnamefont {M.}~\bibnamefont {Agio}},\ }\href
  {\doibase 10.1103/PhysRevLett.101.180404} {\bibfield  {journal} {\bibinfo
  {journal} {Phys. Rev. Lett.}\ }\textbf {\bibinfo {volume} {101}},\ \bibinfo
  {pages} {180404} (\bibinfo {year} {2008})}\BibitemShut {NoStop}%
\bibitem [{\citenamefont {Hucul}\ \emph {et~al.}(2015)\citenamefont {Hucul},
  \citenamefont {Inlek}, \citenamefont {Vittorini}, \citenamefont {Crocker},
  \citenamefont {Debnath}, \citenamefont {Clark},\ and\ \citenamefont
  {Monroe}}]{Hucul2015}%
  \BibitemOpen
  \bibfield  {author} {\bibinfo {author} {\bibfnamefont {D.}~\bibnamefont
  {Hucul}}, \bibinfo {author} {\bibfnamefont {I.~V.}\ \bibnamefont {Inlek}},
  \bibinfo {author} {\bibfnamefont {G.}~\bibnamefont {Vittorini}}, \bibinfo
  {author} {\bibfnamefont {C.}~\bibnamefont {Crocker}}, \bibinfo {author}
  {\bibfnamefont {S.}~\bibnamefont {Debnath}}, \bibinfo {author} {\bibfnamefont
  {S.~M.}\ \bibnamefont {Clark}}, \ and\ \bibinfo {author} {\bibfnamefont
  {C.}~\bibnamefont {Monroe}},\ }\href {http://dx.doi.org/10.1038/nphys3150}
  {\bibfield  {journal} {\bibinfo  {journal} {Nat Phys}\ }\textbf {\bibinfo
  {volume} {11}},\ \bibinfo {pages} {37} (\bibinfo {year} {2015})}\BibitemShut
  {NoStop}%
\bibitem [{\citenamefont {Ghadimi}\ \emph {et~al.}(2016)\citenamefont
  {Ghadimi}, \citenamefont {Blums}, \citenamefont {Norton}, \citenamefont
  {Fisher}, \citenamefont {Connell}, \citenamefont {Amini}, \citenamefont
  {Volin}, \citenamefont {Hayden}, \citenamefont {Pai}, \citenamefont
  {Kielpinski}, \citenamefont {Lobino},\ and\ \citenamefont
  {Streed}}]{Ghadimi2016}%
  \BibitemOpen
  \bibfield  {author} {\bibinfo {author} {\bibfnamefont {M.}~\bibnamefont
  {Ghadimi}}, \bibinfo {author} {\bibfnamefont {V.}~\bibnamefont {Blums}},
  \bibinfo {author} {\bibfnamefont {B.~G.}\ \bibnamefont {Norton}}, \bibinfo
  {author} {\bibfnamefont {P.~M.}\ \bibnamefont {Fisher}}, \bibinfo {author}
  {\bibfnamefont {S.~C.}\ \bibnamefont {Connell}}, \bibinfo {author}
  {\bibfnamefont {J.~M.}\ \bibnamefont {Amini}}, \bibinfo {author}
  {\bibfnamefont {C.}~\bibnamefont {Volin}}, \bibinfo {author} {\bibfnamefont
  {H.}~\bibnamefont {Hayden}}, \bibinfo {author} {\bibfnamefont {C.~S.}\
  \bibnamefont {Pai}}, \bibinfo {author} {\bibfnamefont {D.}~\bibnamefont
  {Kielpinski}}, \bibinfo {author} {\bibfnamefont {M.}~\bibnamefont {Lobino}},
  \ and\ \bibinfo {author} {\bibfnamefont {E.}~\bibnamefont {Streed}},\
  }\href@noop {} {\bibfield  {journal} {\bibinfo  {journal} {arxiv.org}\ ,\
  \bibinfo {pages} {arXiv:1607.00100}} (\bibinfo {year} {2016})}\BibitemShut
  {NoStop}%
\bibitem [{\citenamefont {Teo}\ and\ \citenamefont {Scarani}(2011)}]{Teo2011}%
  \BibitemOpen
  \bibfield  {author} {\bibinfo {author} {\bibfnamefont {C.}~\bibnamefont
  {Teo}}\ and\ \bibinfo {author} {\bibfnamefont {V.}~\bibnamefont {Scarani}},\
  }\href {\doibase http://dx.doi.org/10.1016/j.optcom.2011.05.065} {\bibfield
  {journal} {\bibinfo  {journal} {Optics Communications}\ }\textbf {\bibinfo
  {volume} {284}},\ \bibinfo {pages} {4485 } (\bibinfo {year}
  {2011})}\BibitemShut {NoStop}%
\bibitem [{\citenamefont {Tuchendler}\ \emph {et~al.}(2008)\citenamefont
  {Tuchendler}, \citenamefont {Lance}, \citenamefont {Browaeys}, \citenamefont
  {Sortais},\ and\ \citenamefont {Grangier}}]{PhysRevA.78.033425}%
  \BibitemOpen
  \bibfield  {author} {\bibinfo {author} {\bibfnamefont {C.}~\bibnamefont
  {Tuchendler}}, \bibinfo {author} {\bibfnamefont {A.~M.}\ \bibnamefont
  {Lance}}, \bibinfo {author} {\bibfnamefont {A.}~\bibnamefont {Browaeys}},
  \bibinfo {author} {\bibfnamefont {Y.~R.~P.}\ \bibnamefont {Sortais}}, \ and\
  \bibinfo {author} {\bibfnamefont {P.}~\bibnamefont {Grangier}},\ }\href
  {\doibase 10.1103/PhysRevA.78.033425} {\bibfield  {journal} {\bibinfo
  {journal} {Phys. Rev. A}\ }\textbf {\bibinfo {volume} {78}},\ \bibinfo
  {pages} {033425} (\bibinfo {year} {2008})}\BibitemShut {NoStop}%
\bibitem [{\citenamefont {Kaufman}\ \emph {et~al.}(2012)\citenamefont
  {Kaufman}, \citenamefont {Lester},\ and\ \citenamefont
  {Regal}}]{Kaufman2012}%
  \BibitemOpen
  \bibfield  {author} {\bibinfo {author} {\bibfnamefont {A.~M.}\ \bibnamefont
  {Kaufman}}, \bibinfo {author} {\bibfnamefont {B.~J.}\ \bibnamefont {Lester}},
  \ and\ \bibinfo {author} {\bibfnamefont {C.~A.}\ \bibnamefont {Regal}},\
  }\href {\doibase 10.1103/PhysRevX.2.041014} {\bibfield  {journal} {\bibinfo
  {journal} {Phys. Rev. X}\ }\textbf {\bibinfo {volume} {2}},\ \bibinfo {pages}
  {041014} (\bibinfo {year} {2012})}\BibitemShut {NoStop}%
\bibitem [{\citenamefont {Thompson}\ \emph {et~al.}(2013)\citenamefont
  {Thompson}, \citenamefont {Tiecke}, \citenamefont {Zibrov}, \citenamefont
  {Vuleti\ifmmode~\acute{c}\else \'{c}\fi{}},\ and\ \citenamefont
  {Lukin}}]{Thompson2013}%
  \BibitemOpen
  \bibfield  {author} {\bibinfo {author} {\bibfnamefont {J.~D.}\ \bibnamefont
  {Thompson}}, \bibinfo {author} {\bibfnamefont {T.~G.}\ \bibnamefont
  {Tiecke}}, \bibinfo {author} {\bibfnamefont {A.~S.}\ \bibnamefont {Zibrov}},
  \bibinfo {author} {\bibfnamefont {V.}~\bibnamefont
  {Vuleti\ifmmode~\acute{c}\else \'{c}\fi{}}}, \ and\ \bibinfo {author}
  {\bibfnamefont {M.~D.}\ \bibnamefont {Lukin}},\ }\href {\doibase
  10.1103/PhysRevLett.110.133001} {\bibfield  {journal} {\bibinfo  {journal}
  {Phys. Rev. Lett.}\ }\textbf {\bibinfo {volume} {110}},\ \bibinfo {pages}
  {133001} (\bibinfo {year} {2013})}\BibitemShut {NoStop}%
\bibitem [{\citenamefont {Trautmann}\ \emph {et~al.}(2016)\citenamefont
  {Trautmann}, \citenamefont {Alber},\ and\ \citenamefont
  {Leuchs}}]{Trautmann2016}%
  \BibitemOpen
  \bibfield  {author} {\bibinfo {author} {\bibfnamefont {N.}~\bibnamefont
  {Trautmann}}, \bibinfo {author} {\bibfnamefont {G.}~\bibnamefont {Alber}}, \
  and\ \bibinfo {author} {\bibfnamefont {G.}~\bibnamefont {Leuchs}},\ }\href
  {\doibase 10.1103/PhysRevA.94.033832} {\bibfield  {journal} {\bibinfo
  {journal} {Phys. Rev. A}\ }\textbf {\bibinfo {volume} {94}},\ \bibinfo
  {pages} {033832} (\bibinfo {year} {2016})}\BibitemShut {NoStop}%
\end{thebibliography}
%

\end{document}